\documentstyle[12pt]{article}
\begin{document}

\def\ds{\displaystyle}
\def\beq{\begin{equation}}
\def\eeq{\end{equation}}
\def\bea{\begin{eqnarray}}
\def\eea{\end{eqnarray}}
\def\beeq{\begin{eqnarray}}
\def\eeeq{\end{eqnarray}}
\def\ve{\vert}
\def\vel{\left|}
\def\ver{\right|}
\def\nnb{\nonumber}
\def\ga{\left(}
\def\dr{\right)}
\def\aga{\left\{}
\def\adr{\right\}}
\def\lla{\left<}
\def\rra{\right>}
\def\rar{\rightarrow}
\def\nnb{\nonumber}
\def\la{\langle}
\def\ra{\rangle}
\def\ba{\begin{array}}
\def\ea{\end{array}}
\def\tr{\mbox{Tr}}
\def\ssp{{\Sigma^{*+}}}
\def\sso{{\Sigma^{*0}}}
\def\ssm{{\Sigma^{*-}}}
\def\xis0{{\Xi^{*0}}}
\def\xism{{\Xi^{*-}}}
\def\qs{\la \bar s s \ra}
\def\qu{\la \bar u u \ra}
\def\qd{\la \bar d d \ra}
\def\qq{\la \bar q q \ra}
\def\gGgG{\la g^2 G^2 \ra}
\def\q{\gamma_5 \not\!q}
\def\x{\gamma_5 \not\!x}
\def\g5{\gamma_5}
\def\sb{S_Q^{cf}}
\def\sd{S_d^{be}}
\def\su{S_u^{ad}}
\def\ss{S_s^{??}}
\def\sbp{{S}_Q^{'cf}}
\def\sdp{{S}_d^{'be}}
\def\sup{{S}_u^{'ad}}
\def\ssp{{S}_s^{'??}}
\def\sig{\sigma_{\mu \nu} \gamma_5 p^\mu q^\nu}
\def\fo{f_0(\frac{s_0}{M^2})}
\def\ffi{f_1(\frac{s_0}{M^2})}
\def\fii{f_2(\frac{s_0}{M^2})}
\def\O{{\cal O}}
\def\sl{{\Sigma^0 \Lambda}}
\def\es{\!\!\! &=& \!\!\!}
\def\ar{&+& \!\!\!}
\def\ek{&-& \!\!\!}
\def\cp{&\times& \!\!\!}
\def\se{\!\!\! &\simeq& \!\!\!}
\def\kpm{&\pm& \!\!\!}
\def\kmp{&\mp& \!\!\!}


\def\simlt{\stackrel{<}{{}_\sim}}
\def\simgt{\stackrel{>}{{}_\sim}}


\renewcommand{\textfraction}{0.2}    
\renewcommand{\topfraction}{0.8}   

\renewcommand{\bottomfraction}{0.4}   
\renewcommand{\floatpagefraction}{0.8}
\newcommand\mysection{\setcounter{equation}{0}\section}

\def\baeq{\begin{appeq}}     \def\eaeq{\end{appeq}}  
\def\baeeq{\begin{appeeq}}   \def\eaeeq{\end{appeeq}}
\newenvironment{appeq}{\beq}{\eeq}   
\newenvironment{appeeq}{\beeq}{\eeeq}
\def\bAPP#1#2{
 \markright{APPENDIX #1}
 \addcontentsline{toc}{section}{Appendix #1: #2}
 \medskip
 \medskip
 \begin{center}      {\bf\LARGE Appendix #1 :}{\quad\Large\bf #2}
\end{center}
 \renewcommand{\thesection}{#1.\arabic{section}}
\setcounter{equation}{0}
        \renewcommand{\thehran}{#1.\arabic{hran}}
\renewenvironment{appeq}
  {  \renewcommand{\theequation}{#1.\arabic{equation}}
     \beq
  }{\eeq}
\renewenvironment{appeeq}
  {  \renewcommand{\theequation}{#1.\arabic{equation}}
     \beeq
  }{\eeeq}
\nopagebreak \noindent}

\def\eAPP{\renewcommand{\thehran}{\thesection.\arabic{hran}}}

\renewcommand{\theequation}{\arabic{equation}}
\newcounter{hran}
\renewcommand{\thehran}{\thesection.\arabic{hran}}

\def\bmini{\setcounter{hran}{\value{equation}}
\refstepcounter{hran}\setcounter{equation}{0}
\renewcommand{\theequation}{\thehran\alph{equation}}\begin{eqnarray}}
\def\bminiG#1{\setcounter{hran}{\value{equation}}
\refstepcounter{hran}\setcounter{equation}{-1}
\renewcommand{\theequation}{\thehran\alph{equation}}
\refstepcounter{equation}\label{#1}\begin{eqnarray}}


\newskip\humongous \humongous=0pt plus 1000pt minus 1000pt
\def\caja{\mathsurround=0pt}


\title{
         {\Large
                 {\bf
T--odd correlations in $B \rar K^\ast \ell^+ \ell^-$ decay 
beyond standard model
                 }
         }
      }

\author{\vspace{1cm}\\
{\small T. M. Aliev$^a$ \thanks
{e-mail: taliev@metu.edu.tr}\,\,,
A. \"{O}zpineci$^b$ \thanks
{e-mail: ozpineci@ictp.trieste.it}\,\,,
M. Savc{\i}$^a$ \thanks
{e-mail: savci@metu.edu.tr} \,\,,
C. Y\"{u}ce$^a$}\\
{\small a Physics Department, Middle East Technical University, 
06531 Ankara, Turkey}\\
{\small b  The Abdus Salam International Center for Theoretical Physics,
I-34100, Trieste, Italy} }
\date{}

\begin{titlepage}
\maketitle
\thispagestyle{empty}

\begin{abstract}
T--odd correlations as physical observables in the 
$B \rar K^\ast \ell^+ \ell^-$ decay have been
studied using the most general form of the effective Hamiltonian.
It is observed that these quantities are very sensitive to the new 
physics. We estimate the potential of discovery of these quantities 
at future hadron colliders. 
\end{abstract}

~~~PACS numbers: 12.60.--i, 13.30.--a, 14.20.Mr
\end{titlepage}

\section{Introduction}

Rare $B$ decays, induced by flavor changing neutral current (FCNC) 
$b \rar s(d)$ transition, provide potentially the stringiest testing ground 
in the Standard Model (SM) at loop level. Moreover, $b \rar s(d) \ell^+ \ell^-$
decay is also very sensitive to the new physics beyond the SM. New physics
effects manifest themselves in rare $B$ decays in two different ways, either
through new contributions to the Wilson Coefficients existing in the SM or
through new structures in the effective Hamiltonian which are absent in the
SM.

Recently, time--reversal (T) violation has been measured in the $K^0$
system \cite{R4901}. Unfortunately, the origin of T, as well as CP violation
which also has been obtained experimentally in $K^0$ system, remains
unclear. In the SM, both violations come from a weak phase of the
Cabibbo--Kobayashi--Maskawa (CKM) matrix \cite{R4902}. SM predicts also
violation of CP in the $B^0$ system (see for example \cite{R4903}).
The study of CP violation constitutes one of the main research area of the
working $B$ factories \cite{R4904}. These factories have already reported
evidence for the CP violation in the $B$ systems, namely 
$\sin 2\beta = 0.741 \pm 0.067$ \cite{R4905}. In this work we investigate 
T--violating effects in
the $B \rar K^\ast \ell^+ \ell^-$ using the most general form of the
effective Hamiltonian. It should be noted T--violation effects in the 
$\Lambda_b \rar \Lambda \ell^+ \ell^-$ and $B \rar K^\ast \ell^+ \ell^-$
decays were studied in the framework of the supersymmetric model in
\cite{R4906} and \cite{R4907} as well as in the $\Lambda_b \rar \Lambda
\ell^+ \ell^-$  decay using the most general form of the effective
Hamiltonian in \cite{R4908}, respectively.

It is known that for a general three--body decay, the triplet spin
correlations $\vec{s}\cdot ({\vec{p}}_i \times {\vec{p}}_j)$ are the T--odd
observables, where $\vec{s}$, ${\vec{p}}_i$ and ${\vec{p}}_j$ are the spin
and final momenta of the final particles. Thus in the $B \rar K^\ast \ell^+
\ell^-$ decay, the T--odd observables can be constructed in two different
ways.

\begin{itemize}
\item either by choosing lepton polarization as the polarization of the final particles,
\item or by choosing polarization of $K^\ast$.
\end{itemize}

The first possibility, i.e., the choice of the lepton polarization in the 
$B \rar K^\ast \ell^+ \ell^-$ decay, was studied in detail in \cite{R4909}.
For this reason in the present work, in investigating the T--violating
effects, we choose the second possibility, namely, we choose $K^\ast$
polarization to represent the polarization of the final state.

The paper is organized as follows. In section
2, using the most general, model independent form of the decay amplitude for
the $b \rar s \ell^+ \ell^-$ transition, we study T violation in the 
$B \rar K^\ast \ell^+ \ell^-$ decay. Section 3 is devoted to the numerical 
analysis and concluding remarks.

\section{Theoretical background}

The matrix element of the $B \rar K^\ast \ell^+ \ell^-$ decay is described
by the $b \rar s \ell^+ \ell^-$ transition at quark level. The decay
amplitude for for the $b \rar s \ell^+ \ell^-$ transition, in a general,
model independent form can be written in the following form
\cite{R4909}--\cite{R4911}
\bea
\label{e1}   
\lefteqn{
{\cal M} = \frac{G \alpha}{\sqrt{2} \pi} V_
{tb}V_{ts}^\ast \Bigg\{
C_{SL} \bar s i \sigma_{\mu\nu} \frac{q^\nu}{q^2} L b \bar\ell \gamma^\mu
\ell +
C_{BR} \bar s i \sigma_{\mu\nu} \frac{q^\nu}{q^2} b \bar\ell \gamma^\mu \ell +
C_{LL}^{tot} \bar s_L \gamma^\mu b_L \bar \ell_L\gamma_\mu \ell_L} \nnb\\
\ar C_{LR}^{tot} \bar s_L \gamma^\mu b_L \bar \ell_R \gamma_\mu \ell_R +
C_{RL} \bar s_R \gamma^\mu b_R \bar \ell_L \gamma_\mu \ell_L +
C_{RR} \bar s_R \gamma^\mu b_R \bar \ell_R \gamma_\mu \ell_R \nnb \\
\ar C_{LRLR} \bar s_L b_R \bar \ell_L \ell_R +
C_{RLLR} \bar s_R b_L \bar \ell_L \ell_R +
C_{LRRL} \bar s_L b_R \bar \ell_R \ell_L +
C_{RLRL} \bar s_R b_L \bar \ell_R \ell_L \Bigg\}~,
\eea
where $L=(1-\gamma_5)/2$ and $R=(1+\gamma_5)/2$ are the chiral operators and
$C_X$ are the coefficients of the four--Fermi interaction. Note that this
form of the decay amplitude is motivated by various extensions of the SM,
such as the two Higgs doublet model and supersymmetric models. The first two of
these coefficients, $C_{SL}$ and $C_{BR}$ describe the penguin contributions
which correspond to $-2 m_s C_7^{eff}$ and $-2 m_b C_7^{eff}$ in the SM,
respectively. The next four terms in Eq. (\ref{e1}) represent the vector
type interactions, of whom the two with the coefficients $C_{LL}^{tot}$ and
$C_{LR}^{tot}$ do exist in the SM in the forms $(C_9^{eff}-C_{10})$ and
$(C_9^{eff}+C_{10})$, respectively, i.e.,
\bea
\label{e2}
C_{LL}^{tot} \es C_9^{eff}- C_{10} + C_{LL}~, \nnb \\
C_{LR}^{tot} \es C_9^{eff}+ C_{10} + C_{LR}~.
\eea
The remaining last four terms describe the scalar type interactions. 

The effective Wilson coefficient $C_9^{eff}$ is given by \cite{R4912,R4913}
\bea
\label{e3}   
C_9^{eff} = C_9(\mu) + Y_{pert} + \frac{3 \pi}{\alpha^2} C^{(0)}
\sum_{V_i=J/\psi,\psi^\prime,\cdots} \kappa_i 
\frac{\Gamma(V_i \rar \ell^+ \ell^-) m_{V_i}}
{m_{V_i}^2 - q^2 - i m_{V_i} \Gamma_{V_i}}~,
\eea
where $C^{(0)}=3 C_1 + C_2 + 3 C_3 + C_4 + 3 C_5 + C_6$, $m_{V_i}$ and
$\Gamma(V_i \rar \ell^+ \ell^-)$ are the masses and the widths of the $\psi$
family, and $Y_{pert}(q^2/m_b^2)$ arises from the one--loop matrix element of the
four--quark operators and can be found in \cite{R4912,R4913}. The last term
in Eq. (\ref{e3}) describes the long distance contribution from the real
intermediate $\bar c c$ states \cite{R4914}. The factor $\kappa_i$ for the
lowest resonances are chosen as $\kappa_{J/\psi}=1.65$ and
$\kappa_{\psi^\prime}=2.36$ (see \cite{R4915}) and for the higher resonances
the average of the $\kappa_{J/\psi}$ and $\kappa_{\psi^\prime}$ have
been used.

Exclusive $B \rar K^\ast \ell^+ \ell^-$ decay is described in terms of
matrix elements of the four quark operators in Eq. (\ref{e1}) over meson
states $B$ and $K^\ast$, which are parametrized in terms of form factors. 
The decay amplitude for the $B \rar K^\ast \ell^+ \ell^-$ decays is found to
be
\bea
\label{e4}
{\cal M} \es \frac{G \alpha}{4 \sqrt{2}\pi} V_{tb}V_{ts}^\ast \Bigg\{
\bar \ell \gamma^\mu (1-\gamma_5) \ell \Big[ -2 {\cal V}_{L_1}
\epsilon_{\mu\nu\lambda\sigma} \varepsilon^{\ast\nu} p^\lambda q^\sigma
- i {\cal V}_{L_2} \varepsilon_\nu^\ast + i {\cal V}_{L_3}
( \varepsilon^\ast q)P_\mu + i {\cal V}_{L_4} (\varepsilon^\ast q) q_\mu
\Big] \nnb \\ 
\ar \bar \ell \gamma^\mu (1+\gamma_5) \ell \Big[ -2 {\cal V}_{R_1}
\epsilon_{\mu\nu\lambda\sigma} \varepsilon^{\ast\nu} p^\lambda q^\sigma 
- i {\cal V}_{R_2} \varepsilon_\nu^\ast + i {\cal V}_{R_3}    
( \varepsilon^\ast q)P_\mu + i {\cal V}_{R_4} (\varepsilon^\ast q) q_\mu  
\Big] \nnb \\
\ar \bar \ell (1-\gamma_5) \ell \Big[i {\cal S}_L (\varepsilon^\ast q)\Big]
+ \bar \ell (1+\gamma_5) \ell \Big[i {\cal S}_R (\varepsilon^\ast q)\Big]
\Bigg\}~,
\eea
where $P=p+p_B$, $q=p_B-p$, and $p$ and $\varepsilon$ are the $K^\ast$ meson
four--momentum and four--polarization vectors, and ${\cal V}_{L_i}$ and
${\cal V}_{R_i}$ are the coefficients of left and right handed leptonic
currents with vector structure, and ${\cal S}_{L,R}$ are the coefficients of
the scalar currents with left and right chirality.  
Definitions of the form factors and functions ${\cal V}_{L_i,R_i}$ can be found
in \cite{R4916}. 

In order to obtain T--odd terms $\epsilon_{\mu\nu\alpha\beta}
q^\mu \varepsilon^{\ast\nu} p_\ell^\alpha P^\beta$, we study the 
$B \rar K^\ast \ell^+ \ell^- \rar (K \pi) \ell^+ \ell^-$ process. The
helicity amplitude $M_\lambda^{\lambda_\ell \bar \lambda_\ell}$ of the 
$B \rar K^\ast \ell^+ \ell^-$ decay can be written as
\bea
\label{e5}
M_{\lambda_i}^{\lambda_\ell \bar \lambda_\ell} \es \sum_{\lambda_{V^\ast}} 
\eta_{\lambda_{V^\ast}} L_{\lambda_{V^\ast}}^{\lambda_\ell \bar
\lambda_\ell} H_{\lambda_{V^\ast}}^{\lambda_i}~,
\eea
where 
\bea
\label{e6} 
L_{\lambda_{V^\ast}}^{\lambda_\ell \bar \lambda_\ell} \es 
\varepsilon_{V^\ast}^\mu \lla \ell^-(p_{\ell^-},\lambda_i) 
\ell^+(p_{\ell^+},\bar \lambda_j) \vel J_\mu^\ell \ver 0 \rra~,\nnb \\
H_{\lambda_{V^\ast}}^{\lambda_i} \es
\varepsilon_{V^\ast}^\mu \lla K^\ast (p,\lambda_i) 
\vel J_\mu^i \ver B(p_B) \rra~,
\eea
where $\varepsilon_{V^\ast}$ is the polarization vector of the virtual
intermediate vector boson $(\gamma$ or $Z)$, satisfying the relation
\bea
- g^{\mu\nu} = \sum_{\lambda_{V^\ast}}\eta_{\lambda_{V^\ast}}
\varepsilon_{\lambda_V}^\mu \varepsilon_{\lambda_{V^\ast}}^\nu~,\nnb
\eea
where the summation is over the helicities $\lambda_{V^\ast} = \pm 1,0,s$ of
the virtual intermediate vector boson, with the metric defined as 
$\eta_+ = \eta_0 = -\eta_s = 1$ (see \cite{R4917,R4918}). In Eq. (\ref{e6}),
$J_\mu^\ell$ and $J_\mu^i$ represent the leptonic and hadronic currents,
respectively. 

Using Eqs. (\ref{e4}--\ref{e6}), we get for the helicity amplitudes
\bea
\label{e7} 
{\cal M}_\pm^{++} \es \sin\theta_\ell A_\pm^{++}~,\nnb \\
{\cal M}_\pm^{+-} \es (-1\pm\cos\theta_\ell) A_\pm^{+-}~,\nnb \\
{\cal M}_\pm^{-+} \es (1\pm\cos\theta_\ell) A_\pm^{-+}~,\nnb \\
{\cal M}_\pm^{--} \es \sin\theta_\ell A_\pm^{--}~,\nnb \\
{\cal M}_0^{++} \es \cos\theta_\ell A_0^{++} + B_0^{++}~,\nnb \\
{\cal M}_0^{+-} \es \sin\theta_\ell A_0^{+-}~,\nnb \\
{\cal M}_0^{-+} \es \sin\theta_\ell A_0^{-+}~,\nnb \\
{\cal M}_0^{--} \es \cos\theta_\ell A_0^{--} + B_0^{--}~,
\eea
where $\theta_\ell$ is the polar angle of position in the rest frame of the
intermediate boson with respect to its helicity axis.
Explicit expressions of the functions $A$ and $B$ are presented in the
Appendix (see also \cite{R4916}).

Using the helicity amplitudes given in Eq. (\ref{e7}), the angular
distribution in $B \rar K^\ast \, (\rar K \pi) \ell^+ \ell^-$ is given by
the following expression
\bea
\label{e8}  
d\Gamma \es \frac{3 G^2 \alpha^2}{2^{17} \pi^6 m_B^3 m_\rho^2 q^2} \vel
V_{tb} V_{ts}^\ast \ver B(K^\ast \rar K \pi) dq^2 d\cos\theta_K
d\cos\theta_\ell d\varphi \nnb \\
\cp \lambda^{1/2}(m_B^2,m_{K^\ast}^2,q^2) 
\lambda^{1/2}(m_{K^\ast}^2,m_K^2,m_\pi^2) 
\lambda^{1/2}(q^2,m_\ell^2,m_\ell^2) \nnb \\
\cp \Big\{2 \cos^2\theta_K \Big[\cos^2\theta_\ell N_1 + \sin^2\theta_\ell N_2 +
2 \cos\theta_\ell \, \mbox{\rm Re}(N_3) + N_4 \Big]\nnb \\
\ar \sin^2\theta_K \Big[\sin^2\theta_\ell N_5 + (1+\cos^2\theta_\ell) N_6 + 2
\cos\theta_\ell N_7 + 2 \sin^2\theta_\ell \sin 2\varphi \, \mbox{\rm Im}(N_8) \nnb \\
\ek 2 \sin^2\theta_\ell \cos 2\varphi \, \mbox{\rm Re}(N_8)\Big]
+ \sqrt{2} \sin 2\theta_K \sin\theta_\ell\cos\varphi\, \mbox{\rm
Re}(\cos\theta_\ell N_9 + N_{10}) \nnb \\
\ek \sqrt{2} \sin 2\theta_K \sin\theta_\ell\sin\varphi\, \mbox{\rm
Im}(\cos\theta_\ell N_{11} + N_{12}) \Big\}~.
\eea
Various angles in Eq. (\ref{e8}) are defined as follows: $\theta_K$ is the
polar angle of the $K$ meson in the rest frame of the $K^\ast$ meson,
measured with respect to the helicity axis, i.e., the outgoing direction of
the $K^\ast$ meson. $\theta_\ell$ is the polar angle of the $\ell^+$ in the
dilepton rest frame, measured with respect to the helicity axis of the
dilepton, and $\varphi$ is the azimuthal angle between the two planes
defined by the momenta of the decay products $K^\ast \rar K\pi$ and 
$V \rar \ell^+ \ell^-$. Also, explicit expressions of the functions $N_i$
are given in the Appendix.

It follows from Eq. (\ref{e8}) that terms with $\sim N_8,~N_{11}$ and $N_{12}$ 
contain imaginary part. If we rewrite Eq. (\ref{e8}) for the SM case, we
immediately see that there are two possible sources for T violation:
\begin{itemize}
\item T violation coming from $\mbox{\rm Im} C_9^{eff} C_7^\ast$,
\item T violation coming from $\mbox{\rm Im} C_{10} C_7^\ast$.
\end{itemize}

In SM only $C_9^{eff}$ has imaginary part (see Eq. (\ref{e3}). Therefore we
can conclude that T--odd observables could be nonzero in the processes
involving strong phases or absorptive parts even without weak CP violating
phase. In this work we explore the possibility of the existence of T
violation due to the new weak CP--violating phases. It follows from Eq.
(\ref{e8}) that, in order to have nonvanishing T violation
\begin{itemize}
\item interactions of new type must exist,
\item contributions of different new Wilson coefficients must have weak
CP--violating phases.
\end{itemize}

In order to discard terms $\sim \mbox{\rm Im} C_9^{eff} C_7^\ast$ which give
rise to T--violation in the SM, we consider the following T--odd observable
\bea
\label{e9}
\lla {\cal O} \rra = \int {\cal O} d\Gamma~,
\eea
where ${\cal O}$ is the T--odd correlation, given by 
\bea
\label{e10}
{\cal O} = \frac{(\vec{p}_B\cdot \vec{p}_K) [\vec{p}_B\cdot (\vec{p}_K\times
\vec{p}_{\ell^+})]}{\vel \vec{p}_B \ver^2 \vel \vec{p}_K \ver^2 
(q p_{\ell^+}/\sqrt{2})}~.
\eea
In the $K^\ast $ rest frame, ${\cal O} = \cos\theta_K \sin\theta_K
\sin\theta_\ell \sin\varphi$. The statistical significance of the T--odd
observable in Eq. (\ref{e8}) is determined by \cite{R4907}, 
\bea
\label{e11}
\varepsilon = \frac{\int {\cal O} d\Gamma}{\sqrt{\int d\Gamma}
\sqrt{\int {\cal O}^2 d\Gamma}}~.
\eea
It should be noted that in in Eq. (\ref{e11}), integration over $q^2$ is
carried out in order to eliminate the $q^2$  dependence of $\varepsilon$.
Our final remark in this section is that, T--odd effects that are related 
with the CP violation and CP violating
asymmetry between the decay rates of $B \rar K^\ast \ell^+ \ell^-$ and
$\bar B \rar \bar K^\ast \ell^+ \ell^-$ are discussed in the second 
reference in \cite{R4910}.
\section{Numerical analysis}

In this section we will study the dependence of the statistical significance
$\varepsilon$ on the new Wilson coefficients. For the $B \rar K^\ast$
transition form factors, which are the main input parameters in
$\varepsilon$, we use the light cone QCD sum rules method prediction
\cite{R4919}--\cite{R4921}. The dependence of the form factors on $q^2$ can be
written in terms of the three parameters as
\bea
\label{e12}
F(q^2) = \frac{F(0)}{\ds 1-a_F (q^2/m_B^2) + b_F (q^2/m_B^2)^2}~.
\eea
The value of the parameters $F_i(0),~a$ and $b$ for various form factors
are presented in table 1.
\begin{table}[h]    
\renewcommand{\arraystretch}{1.5}
\addtolength{\arraycolsep}{3pt}
$$
\begin{array}{|lccc|}
\hline
& F(0) & a_F & b_F \\ \hline
A_1 &
0.34 \pm 0.05 & 0.60 & -0.023 \\
A_2 &
0.28 \pm 0.04 & 1.18 &\phantom{-}0.281 \\  
V &
0.46 \pm 0.07 & 1.55 &\phantom{-}0.575 \\  
T_1 &  
0.19 \pm 0.03 & 1.59 &\phantom{-}0.615 \\  
T_2 &  
0.19 \pm 0.03 & 1.49 & -0.241 \\  
T_3 &  
0.13 \pm 0.02 & 1.20 &\phantom{-}0.008 
\\ \hline
\end{array}
$$
\caption{The $B \rar K^\ast$ transition form factors in a three--parameter
fit. The values of the form factors are taken from \cite{R4920}.}
\renewcommand{\arraystretch}{1}
\addtolength{\arraycolsep}{-3pt}
\end{table}

In further numerical analysis, we use next--to leading logarithmic
approximation results for the values of the Wilson coefficients
$C_7,~C_9^{eff}$ and $C_{10}$ at $\mu=m_b$ \cite{R4912,R4913}. As has
already been noted, in the process under consideration, 
only short distance contributions are taken into account in the Wilson
coefficient $C_9^{eff}$ (see the expression for $C_9^{eff}$ given in Eq.
(\ref{e3})).
The new Wilson coefficients vary in the range $-\vel C_{10}\ver \le C_X \le
\vel C_{10} \ver$. 
The experimental bounds on the branching ratio of the 
$B \rar K^\ast \mu^+ \mu^-$ \cite{R4921}
\footnote{The latest result released by the BaBar 
Collaboration for the branching ratio of the $B \rar K^\ast \ell^+ \ell^-$
decay, is
\bea
{\cal B} (B \rar K^\ast \ell^+ \ell^-) = 
\ga 1.68 ^{+0.68}_{-0.58} \pm 0.18 \dr \times 10^{-6}~.\nnb
\eea
} and $B\rar \mu^+ \mu^-$ 
decays \cite{R4922}
suggest that this is the right order of magnitude range for the vector and
scalar interaction coefficients. The present experimental values on the
branching ratio ${\cal B}(B \rar K \ell^+ \ell^-) = \ga
0.78^{+0.24+0.11}_{-0.24-0.11}\times 10^{-6}\dr$ lead to stronger restrictions
on some of the new Wilson coefficients, namely, $-1.5 \le C_T \le 1.5$,
$-3.3 \le C_{TE} \le 2.6$, $-2 \le C_{LL};C_{RL} \le 2.3$ while for all
remaining coefficients $-4 \le C_X \le 4$. Note that if the latest results
for the branching ratio for the $B \rar K^\ast \ell^+ \ell^-$ decay are taken
into account (see the footnote below), the allowed regions of the new
coefficients are $-2.5 \le C_{LL} \le 0$, $0 \le C_{RL} \le 4$ and all
remaining coefficients vary in the region $-4 \le C_X \le 4$.
As has already been noted, in order to
obtain considerable statistical significance $\varepsilon$, the new Wilson
coefficients must have new weak phase. For simplicity we assume that all new
Wilson coefficients have a common weak phase $\phi$. The dependence of the
$\varepsilon$ on the Wilson coefficients $C_{LL},~C_{LR},~C_{RL}$ and
$C_{RR}$ and on the weak phase $\phi$ for the $B \rar K^\ast \mu^+ \mu^-$
decay is presented in Figs. (1)--(4). Note that the dependence of
$\varepsilon$  on the Wilson coefficients for scalar interactions for the $B
\rar K^\ast \mu^+ \mu^-$ decay is not presented since for all their values
$\vel \varepsilon \ver$ is very small $(\le 0.2\%)$. 

From these figures we see that $\varepsilon$ gets its largest value for
$C_{LL}$ about $5\%$, for $C_{LR}$ and $C_{RR}$ about $3\%$ and $C_{RL}$
about $4\%$, for the $B \rar K^\ast \mu^+ \mu^-$ decay.

The situation is quite different from the previous case for the 
$B \rar K^\ast \tau^+ \tau^-$ decay. In this case contributions coming from 
the scalar type interactions are dominant (see Figs. (5)--(8)),
while vector type interactions give negligibly small contributions to 
$\varepsilon$. We observe from these figures that $\varepsilon$ gets    
its maximum value $\sim 4\%$ for $C_{LRLR}$ and $C_{RLLR}$. We also note
that in the present work $C_{BR}$ and $C_{SL}$ are assumed to be identical,
as is the case in the SM, since experimentally measured branching ratio of
$B \rar X_s \gamma$ decay is very close to the SM prediction
\cite{R4923}--\cite{R4925}.

Finally we would like to discuss the detectability of $\varepsilon$ in the
experiments. In order to observe this effect at the $n\sigma$ level, the
required number of $B$ mesons are ${\cal N}_B = n^2 / ({\cal B} \varepsilon^2)$. If
the branching ratio takes on the following values
\bea
{\cal B}(B \rar K^\ast \ell^+ \ell^-) =
\left\{ \begin{array}{l} 2.0 \times 10^{-6}~,~~~~\mbox{\rm for $\mu$ mode}, \\ \\
2.0 \times 10^{-7}~,~~~~\mbox{\rm for $\tau$ mode}~,
\end{array} \right.~, \nnb
\eea
then to be able to observe T--violating effects of ${\cal O}$ in the $B \rar
K^\ast \ell^+ \ell^-$ decay at $3\sigma$ level, with $\varepsilon\sim 3\%$, at least 
\bea
{\cal N}_B=      
\left\{ \begin{array}{l} 5 \times 10^{9}~,~~~~\mbox{\rm for $\mu$ mode},
\\ \\
5 \times 10^{10}~,~~~~\mbox{\rm for $\tau$ mode}~,
\end{array} \right.~, \nnb
\eea
$B$ mesons are needed. Since at LHC and BTeV machines $10^{12} b \bar b$
pairs are expected to be produced per year \cite{R4926}, the observation of
T--violating effects in the $B \rar K^\ast \ell^+ \ell^-$ decay is quite
possible.

\newpage

\bAPP{A}{}

In this appendix we present the explicit expressions of the functions $A$,
$B$ and $N_i$ entering into Eqs. (\ref{e7}) and (\ref{e8}).

\baeeq
A^{++}_{\pm} \es \pm \sqrt{2} m_\ell \Big\{
( C_{LL}^{tot} + C_{LR}^{tot} ) H_\pm +
\frac{2}{q^2} \ga C_{BR} G_\pm + C_{SL} g_\pm \dr  + 
(C_{RR} + C_{RL}) h_\pm\Big\}~, \nnb \\
A^{--}_{\pm} \es - A^{++}_{\pm} ~, \nnb \\
A^{+-}_{\pm} \es \sqrt{\frac{q^2}{2}} \Big\{
\Big[ C_{LL}^{tot} (1-v) + C_{LR}^{tot} (1+v) \Big] H_\pm  
+ \Big[ C_{RL} (1-v) + C_{RR} (1+v) \Big] h_\pm \nnb \\ 
\ar \frac{2}{q^2} \ga  C_{BR} G_\pm + C_{SL} g_\pm \dr \Big\}~, \nnb \\
A^{-+}_{\pm}   \es A^{+-}_{\pm} (v \rar -v)~, \nnb \\
A^{++}_{0}   \es 2 m_\ell \Big[( C_{LL}^{tot} + C_{LR}^{tot} ) H_0 +
(C_{RL} + C_{RR}) h_0 + \frac{2}{q^2} \ga  C_{BR} G_0 + C_{SL} g_0
\dr\Big]~,\\
A^{--}_{0}   \es - A^{++}_{0}~, \nnb \\
B^{++}_{0}   \es - 2 m_\ell \Big\{
(C_{LR}^{tot} - C_{LL}^{tot}) H_S^0 +(C_{RR} - C_{RL}) h_S^0 \Big\} \nnb \\
\ek\frac{2}{m_b} q^2 \Big[ (1-v) ( C_{LRLR}- C_{RLLR}) -
(1+v) ( C_{LRRL}- C_{RLRL}) \Big] H_S^0 ~,\nnb \\
B^{--}_{0}   \es B^{++}_{0} (v \rar -v)~, \nnb \\
A^{+-}_{0}   \es -\sqrt{q^2} \Big\{
\Big[ C_{LL}^{tot} (1-v) + C_{LR}^{tot} (1+v) \Big] H_0 +
\Big[ C_{RL} (1-v) + C_{RR} (1+v) \Big] h_0 \nnb \\ 
\ar \frac{2}{q^2} \ga C_{BR} G_0 + C_{SL} g_0 \dr \Big\}~, \nnb \\
A^{-+}_{0}   \es A^{+-}_{0} (v \rar -v)~,\nnb
\eaeeq
where $~v = \sqrt{1-4 m_\ell^2/q^2}~$ is the lepton velocity and 
superscripts denote helicities of the lepton and antilepton and subscripts
correspond to the helicities of the $K^\ast$ meson, and furthermore 

\baeeq
H_\pm \es \pm \lambda^{1/2}(m_B^2,s_M,q^2) \frac{V(q^2)}{m_B+m_{K^\ast}} +
(m_B+m_{K^\ast}) A_1(q^2)~,\nnb \\
H_0 \es \frac{1}{2 \sqrt{s_M q^2}} \Bigg[
- (m_B^2-s_M-q^2) (m_B+m_{K^\ast}) A_1(q^2)  \nnb \\
\ar\lambda(m_B^2,s_M,q^2) \frac{A_2(q^2)}{m_B+m_{K^\ast}} \Bigg]~,\nnb \\
H_S^0 \es \frac{\lambda^{1/2}(m_B^2,s_M,q^2)}{2 \sqrt{s_M q^2}} \Bigg[
- (m_B+m_{K^\ast}) A_1(q^2) +
\frac{A_2(q^2)}{m_B+m_{K^\ast}} (m_B^2-s_M) \nnb \\ 
\ar 2\sqrt{s_M} (A_3-A_0)\Bigg]~, \\
G_\pm \es - 2 \Big[ \pm \lambda^{1/2}(m_B^2,s_M,q^2) T_1(q^2) +
(m_B^2-s_M)T_2(q^2) \Big]~,\nnb \\
G_0 \es \frac{1}{\sqrt{s_M q^2}} 
\Bigg[ (m_B^2-s_M) (m_B^2-s_M-q^2) T_2(q^2)
- \lambda(m_B^2,s_M,q^2) \Bigg( T_2(q^2) \nnb \\
\ar \frac{q^2}{m_B^2-s_M} T_3(q^2) \Bigg) \Bigg]~,\nnb \\
h_\pm \es H_\pm(A_1 \rar -A_1)~, \nnb \\
\label{hS0}
h_S^0 \es H_S^0(A_1 \rar -A_1,~A_2 \rar -A_2)~,\nnb \\
h_0   \es H_0(A_1 \rar - A_1,~A_2 \rar - A_2)~, \nnb \\
g_\pm \es G_\pm (T_2 \rar - T_2) ~, \nnb \\
g_0 \es - G_0~. \nnb
\eaeeq

\baeeq
N_1 \es \vel A_0^{++} \ver^2 + \vel A_0^{--} \ver^2~, \nnb \\
N_2 \es \vel A_0^{+-} \ver^2 + \vel A_0^{-+} \ver^2~, \nnb \\
N_3 \es A_0^{++}\ga  B_0^{++} \dr^\ast +  
        A_0^{--}\ga  B_0^{--} \dr^\ast ~, \nnb \\
N_4 \es \vel B_0^{++} \ver^2 + \vel B_0^{--} \ver^2~, \nnb \\
N_5 \es \vel A_+^{++} \ver^2 + \vel A_-^{++} \ver^2 +
        \vel A_+^{--} \ver^2 + \vel A_-^{--} \ver^2 ~, \nnb \\
N_6 \es \vel A_+^{+-} \ver^2 + \vel A_-^{-+} \ver^2 +
        \vel A_-^{+-} \ver^2 + \vel A_+^{-+} \ver^2 ~, \nnb \\
N_7 \es \vel A_-^{+-} \ver^2 + \vel A_+^{-+} \ver^2 
        - \vel A_+^{+-} \ver^2 - \vel A_-^{-+} \ver^2 ~, \\
N_8 \es A_+^{++}\ga  A_-^{++} \dr^\ast + 
        A_+^{+-}\ga  A_-^{+-} \dr^\ast + 
        A_+^{-+}\ga  A_-^{-+} \dr^\ast + 
        A_+^{--}\ga  A_-^{--} \dr^\ast ~, \nnb \\
N_{9} \es A_0^{++}\ga  A_-^{++} - A_+^{++} \dr^\ast - 
           A_0^{+-}\ga  A_-^{+-} + A_+^{+-} \dr^\ast - 
           A_0^{-+}\ga  A_-^{-+} + A_+^{-+} \dr^\ast \nnb \\
\ar A_0^{--}\ga  A_-^{--} - A_+^{--} \dr^\ast ~, \nnb \\
N_{10} \es B_0^{++}\ga  A_-^{++} - A_+^{++} \dr^\ast + 
           A_0^{+-}\ga  - A_-^{+-} + A_+^{+-} \dr^\ast + 
           A_0^{-+}\ga  A_-^{-+} - A_+^{-+} \dr^\ast \nnb \\   
\ar B_0^{--}\ga  A_-^{--} - A_+^{--} \dr^\ast ~, \nnb \\
N_{11} \es N_{9} \ga A_+^{++} \rar - A_+^{++},~A_+^{+-} \rar - A_+^{+-},~
           A_+^{-+} \rar - A_+^{-+},~A_+^{--} \rar - A_+^{--} \dr ~, \nnb \\
N_{12} \es N_{10} \ga A_+^{++} \rar - A_+^{++},~A_+^{+-} \rar - A_+^{+-},~
           A_+^{-+} \rar - A_+^{-+},~A_+^{--} \rar - A_+^{--} \dr ~. \nnb
\eaeeq

\eAPP

\newpage

\section*{Figure captions}
{\bf Fig. (1)} The dependence of the statistical
significance $\varepsilon$ on the new Wilson coefficient $C_{LL}$ 
and on the weak phase $\phi$ for the $B \rar K^\ast \mu^+ \mu^-$ decay.\\\\
{\bf Fig. (2)} The same as in Fig. (1), but for the Wilson coefficient
$C_{LR}$.\\\\
{\bf Fig. (3)} The same as in Fig. (1), but for the Wilson coefficient
$C_{RR}$.\\\\
{\bf Fig. (4)} The same as in Fig. (1), but for the Wilson coefficient
$C_{RL}$.\\\\
{\bf Fig. (5)}  The dependence of the statistical
significance $\varepsilon$ on the new Wilson coefficient $C_{LRRL}$ 
and on the weak phase $\phi$ for the $B \rar K^\ast \tau^+ \tau^-$ decay.\\\\
{\bf Fig. (6)} The same as in Fig. (5), but for the Wilson coefficient  
$C_{LRLR}$.\\\\
{\bf Fig. (7)} The same as in Fig. (5), but for the Wilson coefficient
$C_{RLRL}$.\\\\
{\bf Fig. (8)} The same as in Fig. (5), but for the Wilson coefficient  
$C_{RLLR}$.

\newpage

\begin{figure}  
\vskip 0 cm   
    \includegraphics{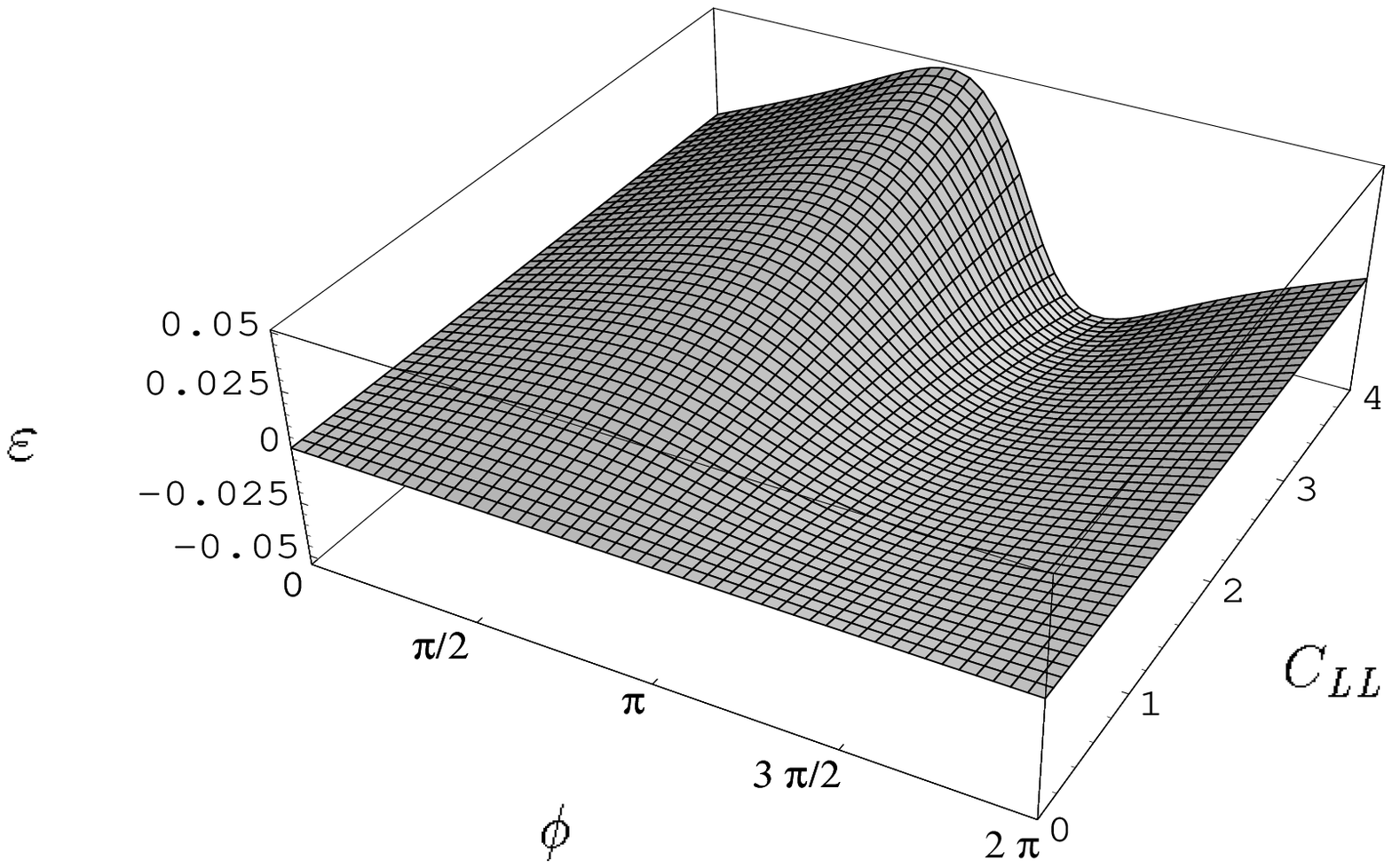}
\vskip 7.0cm     
\caption{}
\end{figure}

\begin{figure}
\vskip 1. cm
    \includegraphics{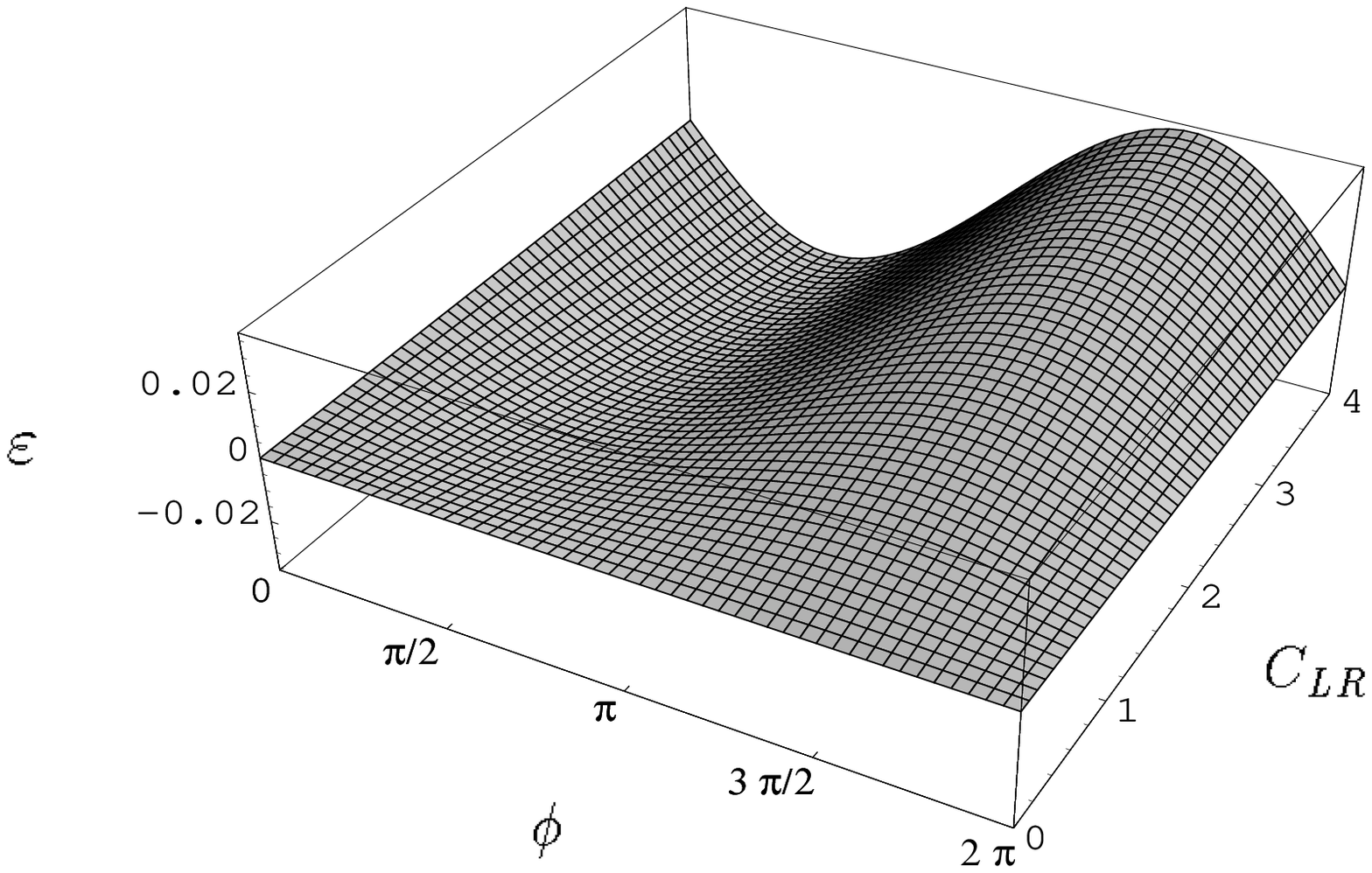}
\vskip 8.0 cm
\caption{}
\end{figure}

\begin{figure}  
\vskip 1.5 cm   
    \includegraphics{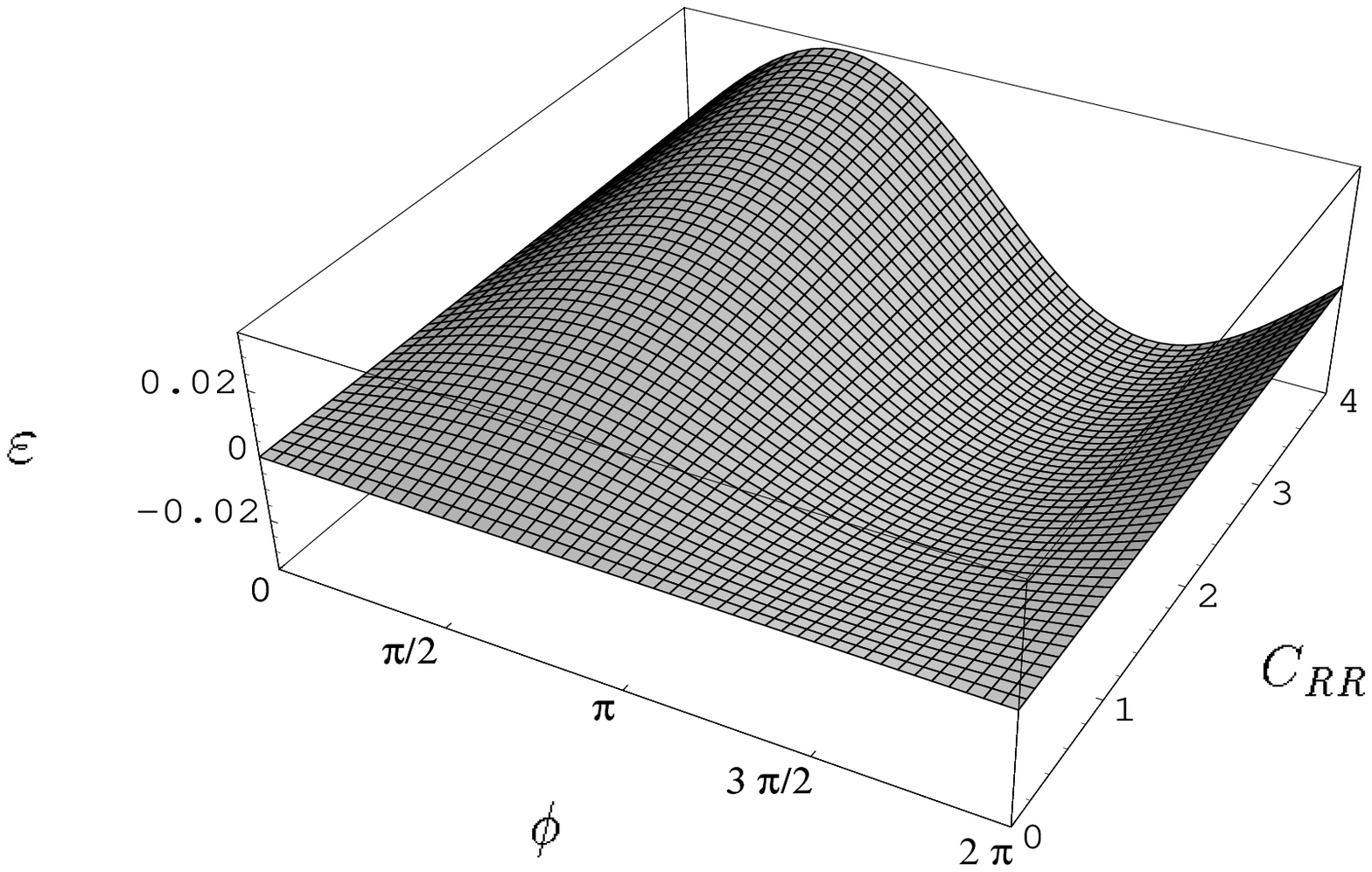}
\vskip 7.0cm     
\caption{}
\end{figure}

\begin{figure}
\vskip 1. cm
    \includegraphics{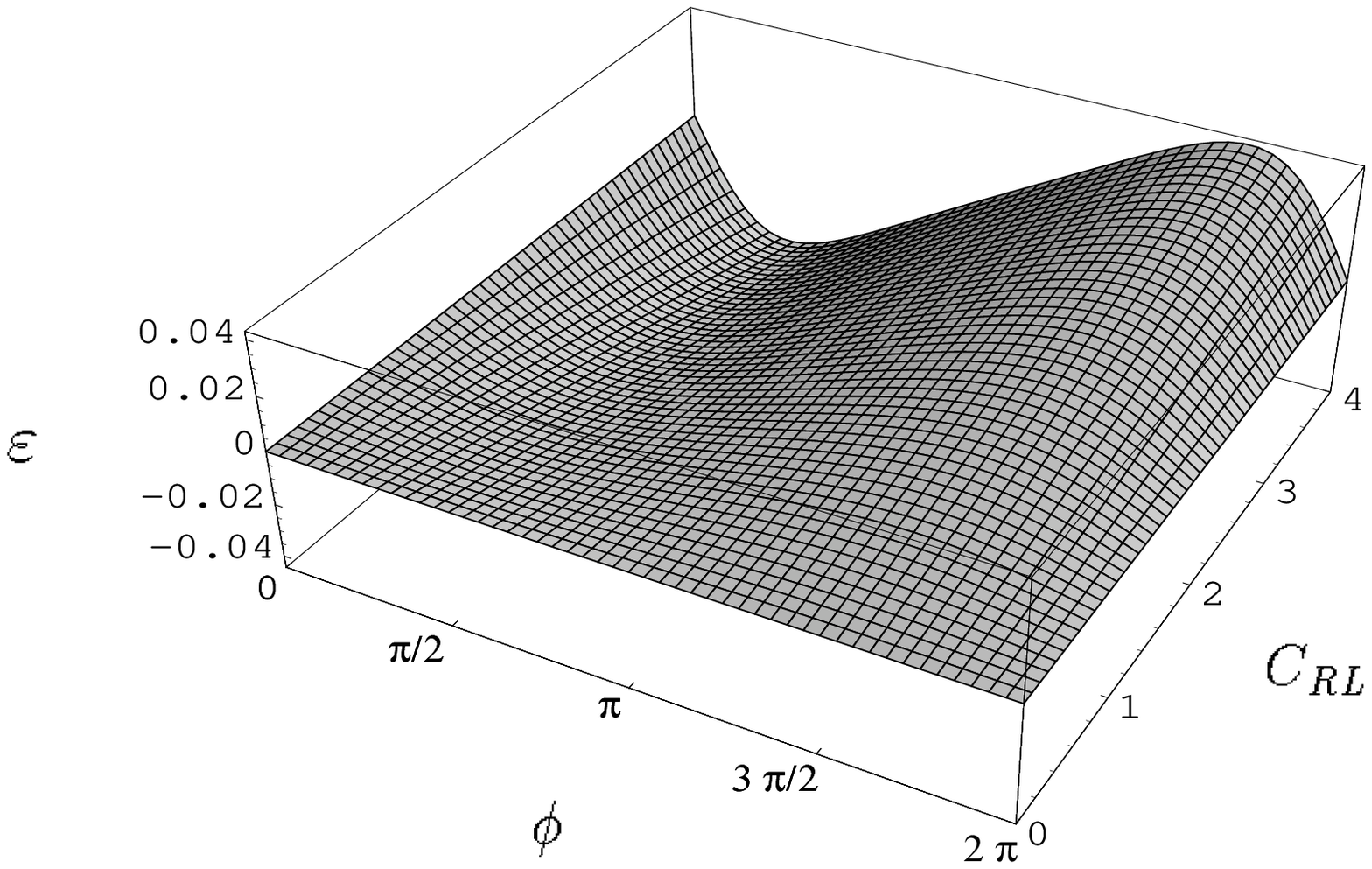}
\vskip 8.0 cm
\caption{}
\end{figure}

\begin{figure}  
\vskip 1.5 cm   
    \includegraphics{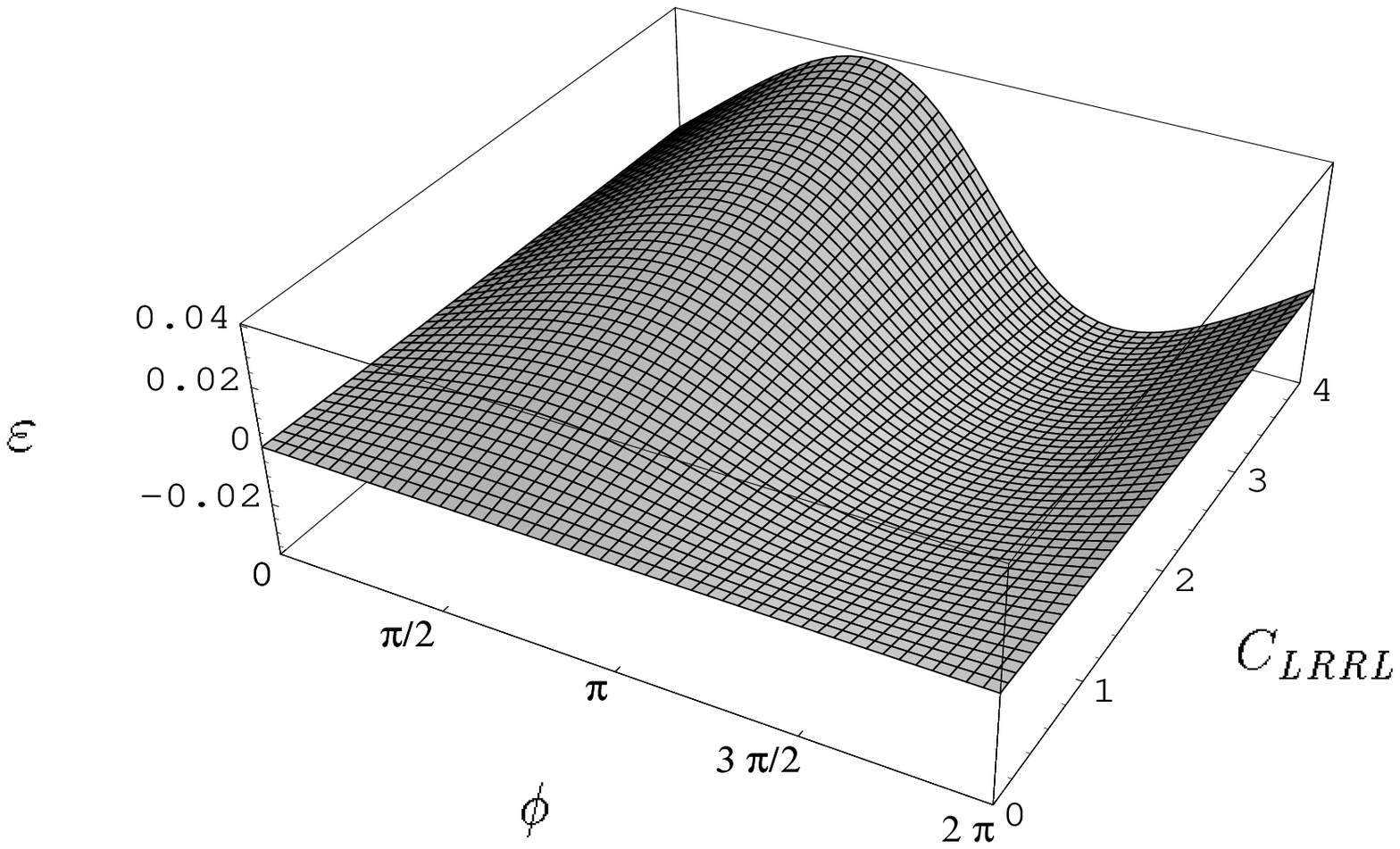}
\vskip 7.0cm     
\caption{}
\end{figure}

\begin{figure}
\vskip 1. cm
    \includegraphics{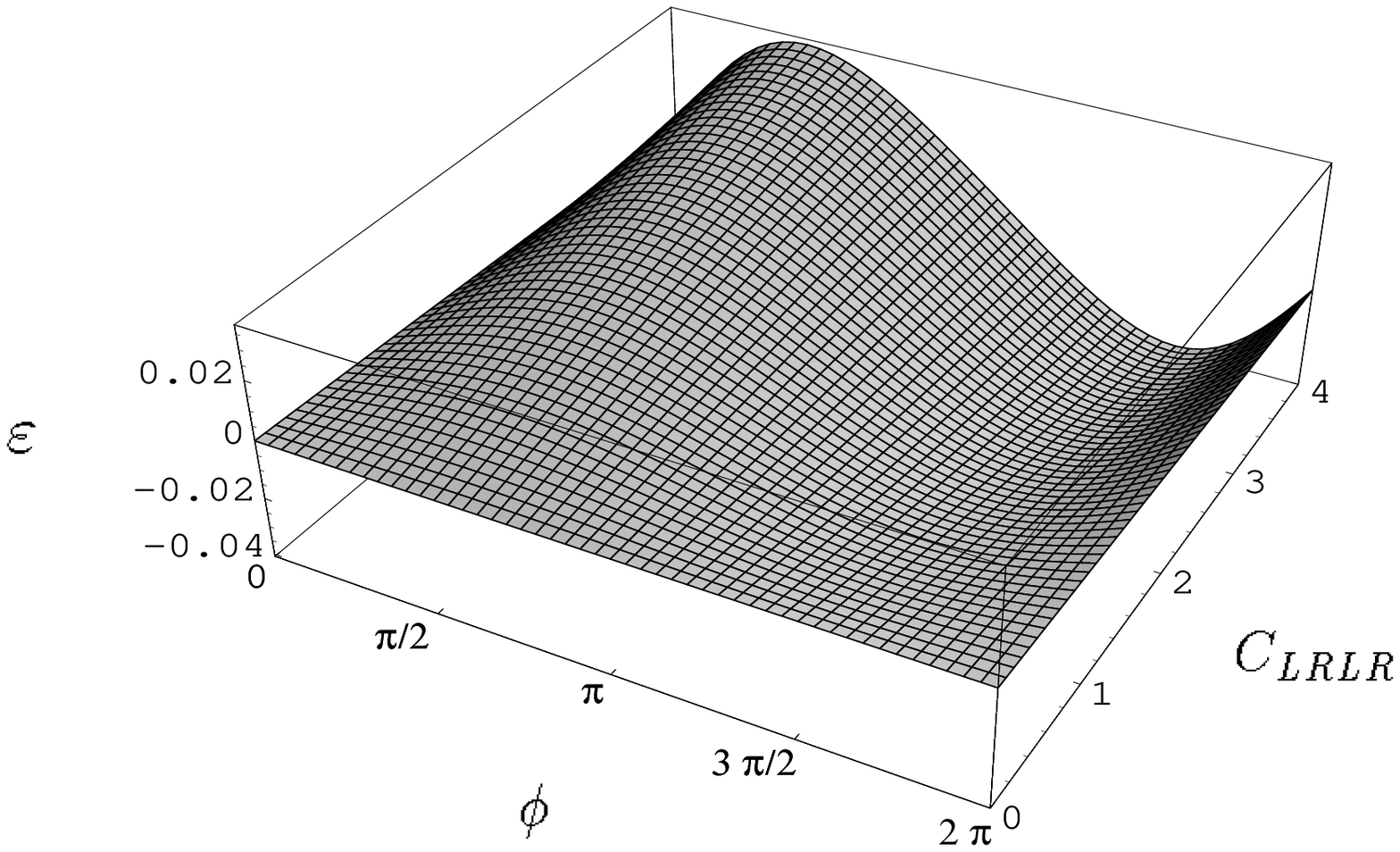}
\vskip 8.0 cm
\caption{}
\end{figure}

\begin{figure}  
\vskip 1.5 cm   
    \includegraphics{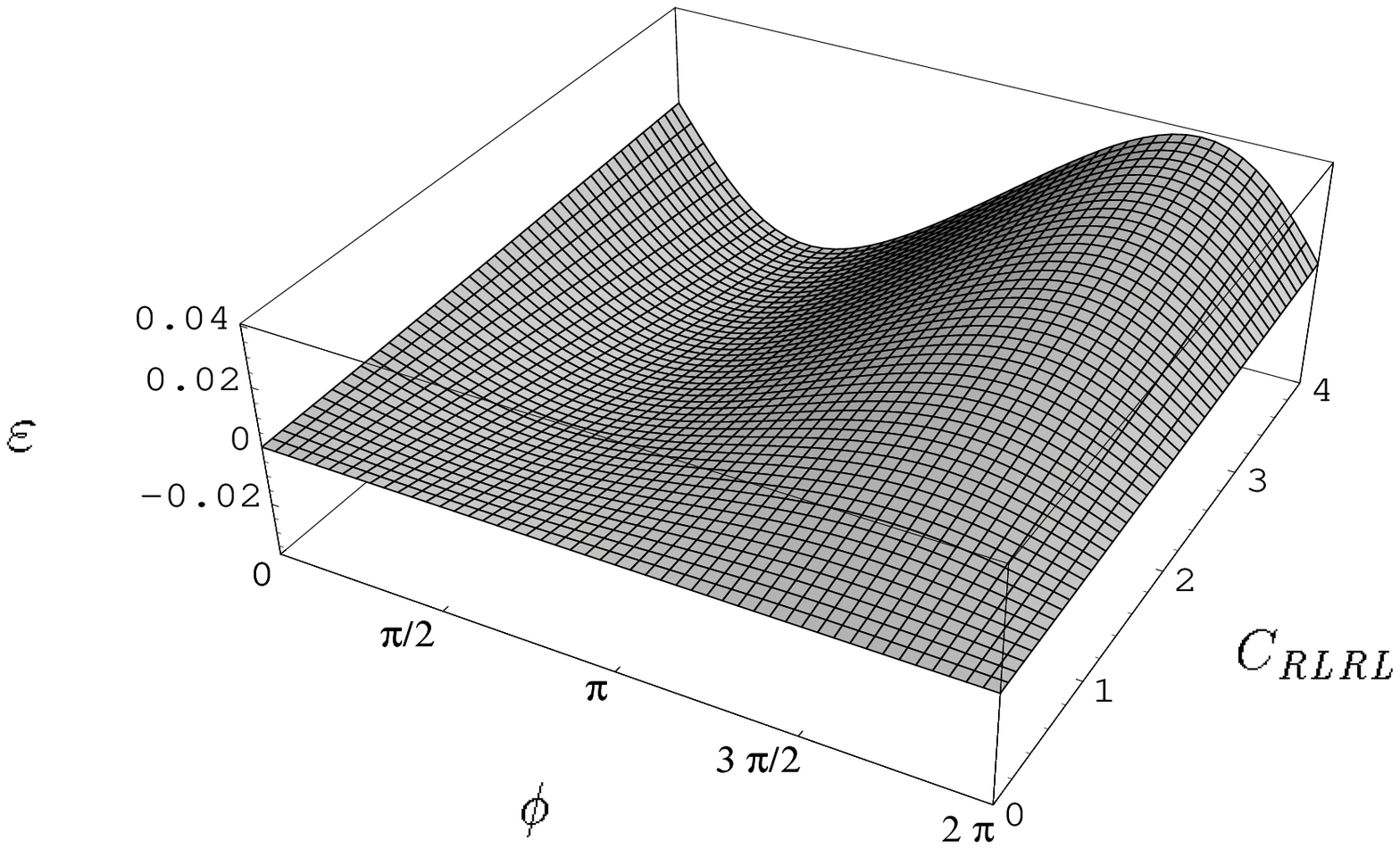}
\vskip 7.0cm     
\caption{}
\end{figure}

\begin{figure}  
\vskip 1.5 cm   
    \includegraphics{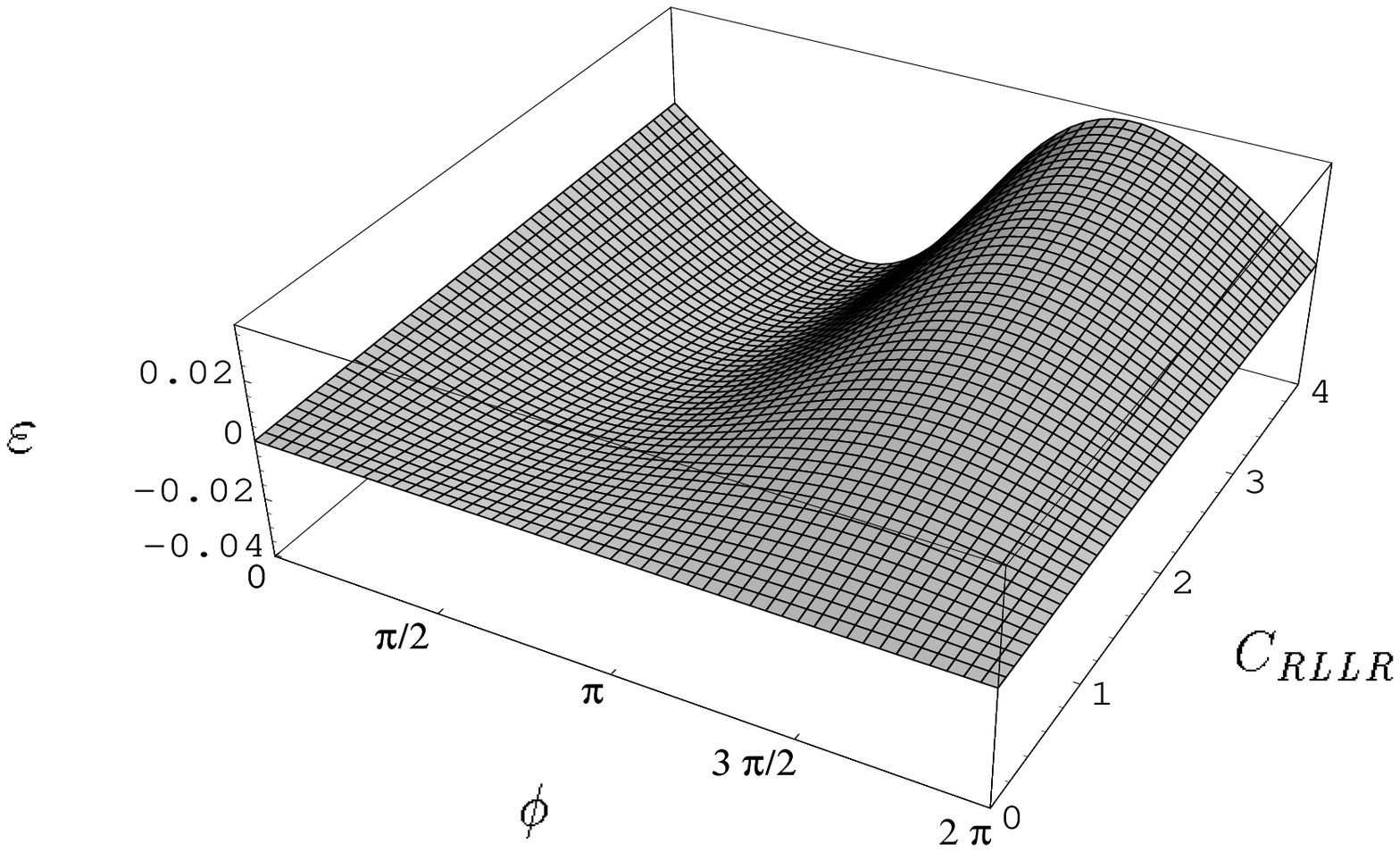}
\vskip 7.0cm     
\caption{}
\end{figure}


\newpage

\begin{thebibliography}{99}

\bibitem{R4901} CPLEAR Collaboration, A. Angelopoulos {\it et. al}
{\it Phys. Lett.} {\bf B444} (1998) 43.

\bibitem{R4902} N. Cabibbo,
{\it Phys. Rev. Lett.} {\bf 10} (1963) 531;\\
M. Kobayashi and T. Maskawa, 
{\it Prog. Theor. Phys.} {\bf 49} (1973) 652.

\bibitem{R4903} C. Caso {\it et. al},
{\it Eur. J. Phys.} {\bf 15} (2000) 1.

\bibitem{R4904} The BaBar Physics book, Eds: P. F. Harrison, H. R. Quinn,
SLAC report (1998) 504;\\
BELLE Collaboration, E. Drebys {\it et. al}, 
{\it Nucl. Inst. Methods} {\bf A446} (2000) 89.

\bibitem{R4905} B. Aubert {\it et. al}, BaBar Collaboration,
hep--ex/0207042 (2002).

\bibitem{R4906} Chuan--Hung Chen, C. Q. Geng and J. N. Ng,
{\it Phys. Rev.} {\bf D65} (2002) 091502.

\bibitem{R4907} Chuan--Hung Chen, C. Q. Geng,
{\it Phys. Rev.} {\bf D66} (2002) 014007.

\bibitem{R4908} T. M. Aliev, A. \"{O}zpineci, M. Savc{\i} and C. Y\"{u}ce,
{\it Phys. Lett.} {\bf B542} (2002) 229.

\bibitem{R4909} T. M. Aliev, M. K. \c{C}akmak, M. Savc{\i},
{\it Nucl. Phys.} {\bf B607} (2001) 305.

\bibitem{R4910} T. M. Aliev, C. S. Kim, Y. G. Kim,
{\it Phys. Rev.} {\bf D62} (2000) 014026;\\
T. M. Aliev, D. Demir, M. Savc{\i},
{\it Phys. Rev.} {\bf D62} (2000) 074016.

\bibitem{R4911} S. Fukae, C. S. Kim, T. Yoshikawa,
{\it Phys. Rev.} {\bf D61} (2000) 074015.

\bibitem{R4912} A. Buras, M. M\"{u}nz,
{\it Phys. Rev.} {\bf D52} (1995) 186.

\bibitem{R4913} M. Misiak,
{\it Nucl. Phys.} {\bf B393} (1993) 23;
Errata {\it ibid} {\bf B439} (1995) 461.

\bibitem{R4914}  N. G. Deshpande, J. Trampetic and K. Panose,
{\it Phys. Rev.} {\bf D39} (1989) 1461;\\
C. S. Lim, T. Morozumi and A. I. Sanda,
{\it Phys. Lett.} {\bf B218} (1989) 343.

\bibitem{R4915} A. Ali, P. Ball, L. T. Handoko and G. Hiller,
{\it Phys. Rev.} {\bf D61} (2000) 074024. 

\bibitem{R4916} T. M. Aliev, M. Savc{\i},
{\it Phys. Rev.} {\bf D62} (2000) 14010;\\
T. M. Aliev, A. \"{O}zpineci, M. Savc{\i},
{\it Nucl. Phys.} {\bf B621} (2002) 479.

\bibitem{R4917} K. Hagiwara, A. D. Martin and M. F. Wade,
{\it Nucl. Phys.} {\bf B327} (1989) 569.

\bibitem{R4918} J. G. K\"{o}rner and G. A. Schuler,
{\it Z. Phys.} {\bf C46} (1990) 93.

\bibitem{R4919} T. M. Aliev, A. \"{O}zpineci, M. Savc{\i},
{\it Phys. Rev.} {\bf D56} (1997) 4260.

\bibitem{R4920} P. Ball and V. M. Braun,
{\it Phys. Rev.} {\bf D58} (1998) 094016. 

\bibitem{R4921} B. Aubert {\it et. al}, BaBar Collaboration,
prep. hep--ex/0207082 (2002). 

\bibitem{R4922} V. Halyo,
prep. hep--ex/0207010 (2002).

\bibitem{R4923} M. S. Alam {\it et. al}, CLEO Collaboration,
{\it Phys. Rev. Lett.} {\bf 74} (1995) 2885;\\
S. Ahmed {\it at. al}, CLEO Collaboration, hep--ex/9908022;\\
D. Cassel , CLEO Collaboration, talk at Lepton Photon Conf. 2001, 
Rome, July 2001.

\bibitem{R4924} R. Barate {\it at. al}, ALEPH Collaboration,
{\it Phys. Lett.} {\bf B429} (1998) 169.

\bibitem{R4925} K. Abe {\it at. al}, BELLE Collaboration,
{\it Phys. Lett.} {\bf B511} (2001) 151.

\bibitem{R4926} N. Harnew, proceedings of heavy flavors 8,
(Southampton, 1999).


\end{thebibliography}
\end{document}